\documentstyle[sprocl2,epsfig,axodraw]{article}
\def\be{\begin{equation}}
\def\ee{\end{equation}}
\def\bea{\begin{eqnarray}}
\def\eea{\end{eqnarray}}
\def\nnb{\nonumber}

\newcommand{\TAI}{\begin{picture}(50,30)(0,10)
\SetScale{0.3}
\SetWidth{3}
\CArc(150,50)(40,0,360)
\end{picture}}

\newcommand{\TBIm}{\begin{picture}(50,30)(0,10)
\SetScale{0.3}
\Line(110,50)(70,50)
\Line(190,50)(230,50)
\SetWidth{3}
\CArc(150,50)(40,0,360)
\end{picture}}

\newcommand{\TBIo}{\begin{picture}(50,30)(0,10)
\SetScale{0.3}
\Line(110,50)(70,50)
\Line(190,50)(230,50)
\CArc(150,50)(40,0,360)
\end{picture}}

\newcommand{\TJIm}{\begin{picture}(50,30)(0,10)
\SetScale{0.3}
\Line(70,50)(230,50)
\SetWidth{3}
\CArc(150,50)(40,0,360)
\end{picture}}

\newcommand{\TJIdm}{\begin{picture}(50,30)(0,10)
\SetScale{0.3}
\Line(70,50)(230,50)
\SetWidth{3}
\CArc(150,50)(40,0,360)
\Vertex(150,90){9}
\end{picture}}

\newcommand{\TJImo}{\begin{picture}(50,30)(0,10)\SetScale{0.3}
\Line(110,50)(190,50)
\SetWidth{3}\CArc(150,50)(40,0,360)
\end{picture}}

\newcommand{\TVI}{\begin{picture}(50,30)(0,10)\SetScale{0.3}
\Line(110,50)(70,50)
\Line(190,50)(230,50)
\CArc(110,10)(40,0,90)
\SetWidth{3}\CArc(150,50)(40,0,360)
\end{picture}}

\newcommand{\TFI}{\begin{picture}(50,30)(0,10)\SetScale{0.3}
\Line(110,50)(70,50)
\Line(190,50)(230,50)
\CArc(150,50)(40,90,270)
\SetWidth{3}
\CArc(150,50)(40,-90,90)
\Line(150,90)(150,10)
\end{picture}}

\newcommand{\TAIcarre}{\begin{picture}(30,30)(0,13)
\SetScale{0.333333}
\SetWidth{3}
\CArc(45,45)(40,0,360)
\end{picture}}

\newcommand{\TBImcarre}{\begin{picture}(50,30)(0,13)
\SetScale{0.3}
\Line(40,50)(0,50)
\Line(120,50)(160,50)
\SetWidth{3}
\CArc(80,50)(40,0,360)
\end{picture}}

\newcommand{\TAIs}{\begin{picture}(90,30)(0,10)
\SetScale{0.3}
\SetWidth{3}
\CArc(150,50)(40,0,360)
\end{picture}}

\newcommand{\TBIms}{\begin{picture}(90,30)(0,10)
\SetScale{0.3}
\Line(110,50)(70,50)
\Line(190,50)(230,50)
\SetWidth{3}
\CArc(150,50)(40,0,360)
\end{picture}}

\newcommand{\TBIos}{\begin{picture}(90,30)(0,10)
\SetScale{0.3}
\Line(110,50)(70,50)
\Line(190,50)(230,50)
\CArc(150,50)(40,0,360)
\end{picture}}

\newcommand{\TJIms}{\begin{picture}(90,30)(0,10)
\SetScale{0.3}
\Line(70,50)(230,50)
\SetWidth{3}
\CArc(150,50)(40,0,360)
\end{picture}}

\newcommand{\TJIdms}{\begin{picture}(90,30)(0,10)
\SetScale{0.3}
\Line(70,50)(230,50)
\SetWidth{3}
\CArc(150,50)(40,0,360)
\Vertex(150,90){9}
\end{picture}}

\newcommand{\TJImos}{\begin{picture}(90,30)(0,10)\SetScale{0.3}
\Line(110,50)(190,50)
\SetWidth{3}\CArc(150,50)(40,0,360)
\end{picture}}

\newcommand{\TVIs}{\begin{picture}(90,30)(0,10)\SetScale{0.3}
\Line(110,50)(70,50)
\Line(190,50)(230,50)
\CArc(110,10)(40,0,90)
\SetWidth{3}\CArc(150,50)(40,0,360)
\end{picture}}

\newcommand{\TFIs}{\begin{picture}(90,30)(0,10)\SetScale{0.3}
\Line(110,50)(70,50)
\Line(190,50)(230,50)
\CArc(150,50)(40,90,270)
\SetWidth{3}
\CArc(150,50)(40,-90,90)
\Line(150,90)(150,10)
\end{picture}}

\begin{document}
\title{Two loop renormalization of the magnetic coupling \\ and non-perturbative sector in hot QCD}
\author{P. Giovannangeli}
\address{Department of Theoretical Physics, University of Bielefeld, Germany\\
E-mail: giovanna@physik.uni-bielefeld.de}

\maketitle

\abstract{The goal of this paper is two-fold. 
The first aim is to present a detailed version of 
the computation of the two-loop renormalization of 
the magnetic coupling in hot QCD. 
The second is to compare with lattice simulations 
the string tension of a spatial Wilson loop using the result 
of our two-loop computation}.

\section{Introduction}

Several breakthroughs have been made in the last few years in the 
computation of high order terms of observables in QCD at high 
temperature. In particular the pressure has been determined perturbatively 
up its last perturbative term namely $g^6 ln(g)$, where $g(T)$ is the 
QCD running coupling~\cite{york}. More than numerical results, 
these computations have given some insights for the knowledge of 
the convergence of perturbation theory in hot QCD. 

Indeed, due to infrared divergences, perturbation theory in hot QCD is limited~\cite{linde}. 
At a given order in the weak coupling expansion, an infinite set of Feynman graphs
will contribute to the thermal mean of an observable. 
This order depends on the observable or thermodynamic quantity that is computed. 
For example, the pressure of the gluon gas can be computed perturbatively up to but not including
the order $g^6$ whereas problems arise at next-to-leading order for the electric Debye mass.
So, despite asymptotic freedom, a non-perturbative part still remains in the hot QCD phase.

Facing such a problem when one does perturbation theory and wanting to apply it in realistic
cases renders mandatory the need for a coherent way to treat the perturbative expansion, 
taking care of infrared divergences. Whatever the approach used to get rid of 
infrared divergences, it is based on the dynamical screening of gluon field in the plasma, 
namely the Debye effect that gives a natural infrared regulator.  

We are going to work in the framework of Dimensional Reduction (D.R.)~\cite{Ginsparg:1980ef,Appelquist:vg}.
It is based on the systematic construction of effective theories 
for each physical scales in the QCD plasma, namely for external momenta of order
$T$, $gT$ and $g^2 T$ . 
This approch combined with $3D$ simulations has provided insights in phase transition 
in the Standard Model~\cite{kajantie} and also allows to compute the perturbative terms 
of the QCD pressure~\cite{york}. 

At really high temperature and large scales, within this method, the QCD plasma is described 
by a magnetostatic Lagrangian whose leading term is a three-dimensional (3D) Yang-Mills 
whose coupling is computed via perturbation theory. 
The important point is that the magnetostatic Lagrangian that describes hot QCD for external momenta of order $g^2 T$, 
contains all the non-perturbative physics in the region where the coupling is small.\\

Our purpose is first to compute
the effective coupling of this 3D Yang-Mills~\cite{moi} by integrating out the
electrostatic Debye scale of order $g T$ and then to use this result in order to investigate, 
thanks to lattice simulations, how our actual knowledge of the magnetostatic part 
is sufficient to encompass the whole non-perturbative high T physics in the case of
the spatial Wilson loop. 
This article is composed as follows:\\
Section~(\ref{sec:DR}) is a brief recall of the principles of D.R..
Section~(\ref{sec:mag})  focuses on the magnetostatic action. 
Section~(\ref{sec:scalarisation})  shows the steps to perform the two-loop computation.
Section~(\ref{sec:transv}) is the body of the computation with all intermediate and final
result. Finally the last section will deal with the applications of magnetostatic physics in 
the QCD plasma, looking at an observable, the so-called Wilson Loop.

\section{Dimensional reduction perturbation theory}\label{sec:DR}

Due to infrared divergences, perturbation theory in hot QCD is limited. That means that 
the analytical computation of the thermal mean of an observable is confronted 
with two main problems.
The first one is to make a coherent perturbation theory until it breaks down.
The second one is to evaluate numerically and with the most precision the contribution of the
non perturbative sector. 
Indeed whatever the observable you will consider, the weak coupling expansion 
will be limited.

For a review of effective theories and dimensional reduction see~\cite{Shaposhnikov:1996th}. 
At equilibrium, a thermodynamic system is described by its partition function:

\begin{equation}
Z(\beta) = Tr_{phys} \exp{(-\beta H)}
\end{equation}

\noindent where $H$ is the Hamiltonian of the system.
In the imaginary time formalism, the latter can be written 
as a path integral:

\begin{equation}
Z(\beta) = \int D\phi D\psi \exp{(- \int_0^{1\over T} \int d^3 x dt {\cal L_E})}
\end{equation}

\noindent with $\phi$ a bosonic field and $\psi$ a fermionic field,
with the boundary conditions $\phi(x, {1\over T})=\phi(x, 0)$ and $\psi(x, {1\over T})= -
\psi(x, 0)$. So, technically, one is dealing with quantum field theory on a cylinder with a
compact time direction, as in Kaluza-Klein theories. The periodicity of the time dimension is $1 \over T$.
So, for large $T$, i.e. larger than the typical momentum scale that one considers, one
expects to have an effective three dimensional theory. Due to time periodicity, 
as for massless bosonic fields, the euclidean propagator reads:

\begin{equation}
{1\over {p^2}} = {1\over \vec p^2 + (2 \pi n T)^2}.
\end{equation}

\noindent The massless bosons have a mass $2 \pi n T$ that is going to 
regularize the infrared.
But for $n= 0$, i.e static modes, infrared problems will still be there. 
Let us be more precise.

In the case of QCD, we have:

\begin{equation}
{\cal L}_{QCD}={1\over 2}Tr F_{\mu\nu}F_{\mu\nu}+\sum_{i=1}^{N_f}\bar
\psi_i(\gamma_{\mu}D_{\mu}+m_i)\psi_i.
\label{lagQCD}
\end{equation}

\noindent As quarks have no zero Matsubara frequency, they always appear as heavy modes.

But what is going to happen for the $A_\mu$ static modes?
Part of the problem will be solved because for electric static mode $A_0$, there is 
Debye screening that is going to regularize the infrared. Indeed, like in classical plasmas, the 
chromoelectric field is screened in such a state of matter. The electrostatic potential  
will behave as Yukawa potential, and large distance contribution will be suppressed thanks 
to the exponential fall-off of such a potential. We describe below in a more quantitative way what is going to happen.

In fact there will be different physical scales in the plasma, 
to which different effective theories will correspond. Let $p$ be the value of the external momentum in the plasma.

\begin{figure}
\begin{center}
\includegraphics{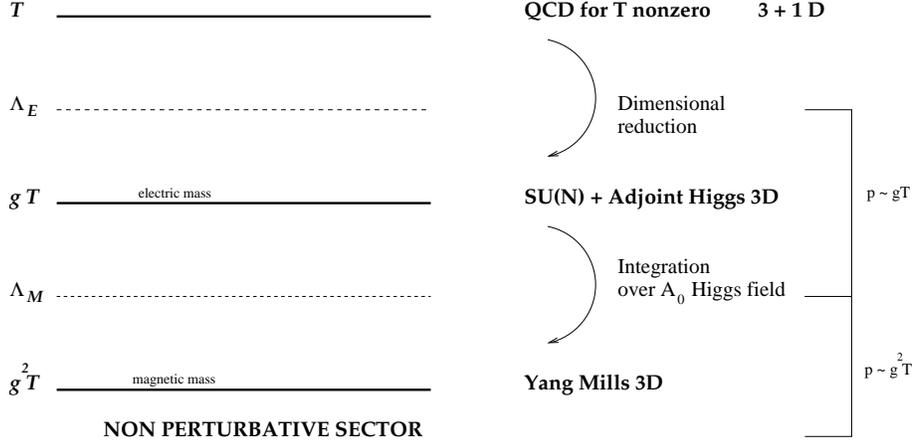}
\caption{Dimensional reduction at work. Integrating out each physical scale of the plasma will lead to a new effective theory.}
\label{fig:kklein}
\end{center}
\end{figure}

\begin{itemize}
\item for $p \sim T$, the theory is QCD for hard modes ( with $n$ nonzero)
\item  for $p \sim gT$, we have integrated out hard modes and obtained an effective  
3D action for electrostatic and magnetostatic modes.
\item for $p \sim g^2 T$, we have integrated out electrostatic modes as they have a Debye mass $\sim gT$. This leaves us with a magnetostatic Lagrangian.
\end{itemize}
\noindent As we integrated out non-static modes, the two effectives theories are three-dimensional. This
dimensional reduction is the same process as in Kaluza-Klein theories.
The procedure is schematically described in Fig.~\ref{fig:kklein}.


In a more precise way, we have:

\begin{itemize}
\item for the hard modes: eq.~(\ref{lagQCD}).
\item for the soft modes, after integrating out the heavy modes, we have
an $SU(N)$ Yang-Mills + adjoint Higgs effective theory.
Its Lagrangian reads:
\begin{eqnarray}
{\cal{L}}_{E} & = & Tr(\vec D(A)A_0)^2+m_E^2TrA_0^2+\lambda_E(Tr(A_0^2))^2+ {} \nonumber\\
              & + &  \bar\lambda_E \big(Tr(A_0)^4-{1\over 2}(TrA_0^2)^2\big)+{1\over 2}Tr F_{ij}^2+\delta{\cal{L}}_E.
\label{estat}
\end{eqnarray}

with as parameters $g_E( g, T)$, $m_E (g, T)$, etc ...
and $\delta{\cal{L}}_E$ represent higher orders operators.

\item for the ultra-soft modes, the effective theory is a magnetostatic Yang-Mills:

\begin{eqnarray}
{\cal{L}}_{M} & = & {1\over 2}Tr F_{ij}^2+\delta{\cal{L}}_M.
\label{estat}
\end{eqnarray}

with as parameter $g_M( g_E, m_E,$ etc..)
and $\delta{\cal{L}}_M$ represent higher orders operators.
\end{itemize}

All theses Lagrangians are deduced from each other via perturbation theory. That
means that they are known to a given order and practically one uses the truncated
Lagrangians, for example ${\cal{L}}_{M}-\delta{\cal{L}}_M$, to do some
computation.
Of course it is important to know which orders are not taken into account. We
will discuss this in the next section.

The technical determination of the magnetostic Lagrangians will be discussed in the next chapter.
We can already make some remarks: ${\cal{L}}_{M}$ is non-perturbative. The theory
has a dimensionful coupling constant
$g_M^2 \sim g^2 T$. So if one wants to use perturbation theory, the dimensionless expansion
parameter is ${g_M^2 \over|\vec p|}$ which is order $1$ for $ |\vec p|\sim g^2 T$!

Finally, once one has computed the parameters, one has to go to the lattice and use ${\cal{L}}_{M}$
to determine magnetic properties of the quark-gluon plasma.

\section{Magnetostatic action in hot QCD}\label{sec:mag}

As previously explained we are going to focus our attention on the magnetostatic
action  ${\cal{L}}_{M}$. The reason is the following: 
the non-perturbative terms in the weak coupling expansion of a thermodymamic quantity 
or an observable come from the magnetostatic Lagrangian. 
So its study gives us some clues to study 
numerically and with precision the convergence of the perturbative series at high
$T$. 
Namely, at which degree of precision the magnetostatic Lagrangian has to be
known in order to include the major part of non-perturbative effects? The goal of this first
part is to determine this Lagrangian up to relative order $g^2$. 
First we are going to recall explicitely how this magnetostatic Lagrangian is constructed by
integrating out the Debye scale in the electrostatic action. Then, 
as the electrostatic Debye scale is integrated out via perturbation theory, 
we are going to discuss  at which order we make the truncation (i.e. what is the order of $\delta{\cal{L}}_M$)  and show the computation of the magnetostatic coupling $g_M$ at two loop order.

\subsection{From the electrostatic action to the magnetostatic action}

The electrostatic action is determined by integrating out heavy modes in the full QCD
Lagrangian, namely gluons with a non-zero Matsubara frequency and quarks. 
Its domain of validity is for external momenta of order $g T$. Then for momenta of order
$g^2 T$ one has to integrate the Debye scale. 
A well-known shortcut is by making a background expansion 
of the effective potential for the magnetostatic field ~\cite{abbott}.

Consider $S_M$ the full magnetostatic action containing the $3 D$ Yang-Mills and all other
higher order terms and $S_E$ the electrostatic action.
The static fields in the electrostatic Lagrangian 
are the massless magnetostatic $A_i$ field and the 
electrostatic $A_0$ field with mass $m_E$. 
In order to integrate out the Debye scale, one has
to introduce a background field for the magnetostatic field, namely,

\be
\vec A = \vec B + g_E \vec Q 
\ee

\noindent with $\vec B$ the background and $\vec Q$ the quantum fluctuations.

Then formally we can construct the magnetostatic action: 

\begin{equation}
\exp{[- {1\over g_M ^2}S_M(B_i)]} = \int DA_0 DQ_i exp{[-{1\over g_E ^2}
  S_E(A_0, B_i +g_E Q_i)]}
\label{reduc}
\end{equation}

\noindent where the integration is from external momenta of order $g^2 T$ and for diagrams with
at least one $A_0$ in the internal loops.
This effective action is computed via perturbation theory. 
That means that one has to use truncation when some computations (namely lattice 
computations in this case) has to be performed with this magnetostatic action.

\subsection{Truncated magnetostatic Lagrangian and precision}

Now the aim is to see what is the relative error made by using the truncated
magnetostatic Lagrangian.

At this point, it is interesting to recall again that all of the non perturbative part is contained in
the magnetostatic Lagrangian. So if one wants to know more about the structure of the 
non-perturbative terms following a perturbative expansion, for example for the order up to and
including $g^6$ for the pressure, its dominant contribution will certainly come from
 the zero-loop magnetostatic Lagrangian, namely a 3D Yang Mills theory. 
In fact the whole question is to mesure with lattice simulations the weight of this
contribution and also to see if the high order operators  in the magnetostatic Lagrangian will
be important. This will be the aim of the second part of this paper. 

Now, let us take a look at higher order operators. One has to compute the one loop expansion
in eq.~(\ref{reduc}).  It gives two operators, namely $ g_E^2 Tr F_{ij} F_{jk} F_{ki}\over m_E^3$ and $ g_E^2 (D_i F_{ij})^2 \over m_E^3$ ~\cite{moibientot}. What is the order of such terms? For example, for the second term the integration of the hard modes 
have provided that at zero-loop $g_E \sim g^2 T$ and at one-loop
$m_E \sim g T$. So, as in this sector, $D \sim g^2 T$, we conclude that $\delta{\cal{L}}_M$   
is of relative order $g^3$. The same occurs for the first term.

To put it in a nutshell, to compute corrections of order $g$ to the $3D$ magnetostatic Yang-Mills, the one-loop renormalization
of $g_M$ is needed. For the correction of order $g^2$ to the $3D$ magnetostatic Yang-Mills, 
the two loop renormalization of $g_M$ is needed. Finally for correction of order $g^3$ to the
$3D$ magnetostatic Yang-Mills, the three-loop renormalization of $g_M$ but also the first term
in  $\delta{\cal{L}}_M$ are needed.
Now let us go into more details for what is needed for the two-loop renormalization
of $g_M$.

\subsection{The coupling of the magnetostatic Lagrangian}

As we have seen, to determine $\delta{\cal{L}}_M$ at two loop order is to determine its coupling
$g_M$ at two loop.

For this we need to compute the two-point function for the background field with at least 
one $A_0$ into the loop.
We calculate the fluctuations around the background in a path integral:
\begin{equation}
\exp{-{1\over{g_M^2}}S_M(B)}=\int DA_0DQ_i\exp {1\over g_E^2}{\big (-S_E-{1\over{\xi}}Tr(D_iQ_i)^2\big)}.
\label{path}
\end{equation}

\noindent We added a general background gauge term. The resulting action $S_M(B)$ is
gauge invariant to all loop orders and the renormalization of the coupling
is identified from the background field two point function at a momentum $p=O(g^2T)$:

\begin{equation}
 \exp{\big(-{1\over{g_M^2}}S_M(B)\big)}= \exp{\big(-{1\over{g_E^2}}S_M(B)\big)}\big(1+(F_1^{tr}+F_2^{tr}+...)S_M(B)\big).
\label{ident}
\end{equation}

\noindent Here $F_i$ is the sum of all Feynman diagrams  for the two point function of the background field with $i$ loops and reads:

\begin{equation}
F_i=F_i^{tr}~(\delta_{lm}p^2-p_lp_m).
\end{equation}

\noindent This leaves us with the relation, using eq.(\ref{ident}):
\begin{equation}
{1\over{g_M^2}}={1\over{g_E^2}}-F_1^{tr}-F_2^{tr},
\end{equation}

\noindent with

\begin{eqnarray}
g_E^2F_1^{tr} & = & -{1\over{48}}{g_E^2N\over{\pi m_E}}\\
g_E^2F_2^{tr} & = & -{19\over{4608}}({g_E^2N\over{\pi m_E}})^2.
\label{final}
\end{eqnarray}

The next section shows the details of the computation that leads to this result.

\section{General and systematic method to compute diagramms}\label{sec:scalarisation}

The following section aims to recall the systematic methods in order to compute
a high order loop expansion. It deals of course with well known facts but it is
important to really go into detail in each of these steps in order to see 
how it efficiently deals with the problem, which is too big to be treated by hand, but
simple enough to present intermediate steps explicitely.

First one has to fix the gauge for the magnetostatic field in the Lagrangian.
We are going to take the background gauge that insure the gauge independence of
our result.

After gauge fixing, our whole electrostatic Lagrangian reads:

\bea
{\cal{L}}_{E}  &=&  Tr(\vec D(\vec B + g_E \vec Q)A_0)^2+m_E^2TrA_0^2+\lambda_E(Tr(A_0^2))    \nonumber\\
              & + &  \bar\lambda_E \big(Tr(A_0)^4-{1\over2}(TrA_0^2)^2)\\
              & + & {1\over 2}Tr F_{ij}(\vec B + g_E \vec Q)^2+ {1\over \xi}(D_{i}(\vec B)Q_i)^2 \\
              &+& \delta{\cal{L}}_E.
\eea

\noindent Let us recall that the vertices and propagators involving a magnetostatic field
are going to be gauge dependent. 
Here are the Feynman rules with:

\begin{itemize}
\item lines for $A_0^a$
\item wavy lines for $A_i^a$
\item a dot for the background field
\end{itemize}

They reads:


\noindent\begin{picture}(100,50)(0,25)
\Gluon(0,25)(55,25){3}{6}
\Text(-5,25)[r]{$\vec{p}$,a,i}
\Text(60,25)[l]{$-\vec{p}$,b,j }
\end{picture}
$= {\delta_{ab} \over p^2} ( \delta_{ij} - (1-\xi){p_i p_j\over p^2})$

\noindent\begin{picture}(100,50)(0,25)
\Line(0,25)(55,25)
\Text(-5,25)[r]{$\vec{p}$,a}
\Text(60,25)[l]{$-\vec{p}$,b}
\end{picture}
$= {\delta_{ab} \over p^2 + m^2}$


\noindent \begin{picture}(100,50)(0,25)
\Gluon(0,25)(30,25){3}{4}
\Line(30,25)(50,45)
\Line(30,25)(50,05)
\Text(-5,25)[r]{$\vec{p}$,a,i}
\Text(55,45)[l]{$\vec{q}$,b}
\Text(55,05)[l]{$\vec{r}$,c}
\end{picture}
$=-i g f_{abc}(r_i-q_i)$

\noindent \begin{picture}(100,50)(0,25)
\Gluon(0,25)(30,25){3}{4}
\Gluon(30,25)(50,45){3}{4}
\Gluon(30,25)(50,05){3}{4}
\Text(-5,25)[r]{$\vec{p}$,a,i}
\Text(55,45)[l]{$\vec{q}$,b,j}
\Text(55,5)[l]{$\vec{r}$,c,k}
\end{picture}
$=-i g f_{abc}(\delta_{ik}(p_j-r_j)+\delta_{jk}(r_i-q_i)+\delta_{ij}(q_k-p_k))$
 
\noindent\begin{picture}(100,50)(0,25)
\Gluon(30,25)(10,45){3}{4}
\Gluon(30,25)(10,5){3}{4}
\Gluon(30,25)(50,45){3}{4}
\Gluon(30,25)(50,05){3}{4}
\Text(5,45)[r]{a,i}
\Text(5,5)[r]{b,j}
\Text(55,45)[l]{c,k}
\Text(55,5)[l]{d,l}
\end{picture}
$=-g^2(f_{abe}f_{ecd}(\delta_{ik} \delta_{jl}-\delta_{il}\delta_{jk})$
\begin{center}
$+f_{ade}f_{ebc}(\delta_{ij}\delta_{kl}-\delta_{ik}\delta_{jl})$
\end{center}
\begin{center}
$+f_{ace}f_{ebd}(\delta_{ij}\delta_{kl}-\delta_{il}\delta_{jk}))$
\end{center}

\newpage

\noindent\begin{picture}(100,50)(0,25)
\Gluon(30,25)(10,45){3}{4}
\Gluon(30,25)(10,5){3}{4}
\Line(30,25)(50,45)
\Line(30,25)(50,05)
\Text(5,45)[r]{a,i}
\Text(5,5)[r]{b,j}
\Text(55,45)[l]{c}
\Text(55,5)[l]{d}
\end{picture}
$=-g^2(f_{ade}f_{ebc}+f_{ace}f_{ebd})\delta_{ij}$


\noindent\begin{picture}(100,50)(0,25)
\Gluon(0,25)(30,25){3}{4}
\Line(30,25)(50,45)
\Line(30,25)(50,05)
\Vertex(0,25){3}
\Text(-5,25)[r]{$\vec{p}$,a,i}
\Text(55,45)[l]{$\vec{q}$,b}
\Text(55,05)[l]{$\vec{r}$,c}
\end{picture}
$=-i g f_{abc}(r_i-q_i)$  

\noindent\begin{picture}(100,50)(0,25)
\Gluon(0,25)(30,25){3}{4}
\Gluon(30,25)(50,45){3}{4}
\Gluon(30,25)(50,05){3}{4}
\Vertex(0,25){3}
\Text(-5,25)[r]{$\vec{p}$,a,i}
\Text(55,45)[l]{$\vec{q}$,b,j}
\Text(55,5)[l]{$\vec{r}$,c,k}
\end{picture}
$=-i g f_{abc}(\delta_{ik}(p_j-r_j-{q_j \over \xi})$
\begin{center}
$+\delta_{jk}(r_i-q_i)$
\end{center}
\begin{center}
$+\delta_{ij}(q_k-p_k+{r_k\over \xi}))$
\end{center}

\noindent\begin{picture}(100,50)(0,25)
\Gluon(30,25)(10,45){3}{4}
\Gluon(30,25)(10,5){3}{4}
\Gluon(30,25)(50,45){3}{4}
\Gluon(30,25)(50,05){3}{4}
\Vertex(10,5){3}
\Text(5,45)[r]{a,i}
\Text(5,5)[r]{b,j}
\Text(55,45)[l]{c,k}
\Text(55,5)[l]{d,l}
\end{picture}
$=-g^2(f_{abe}f_{ecd}(\delta_{ik} \delta_{jl}-\delta_{il}\delta_{jk})$
\begin{center}
$+f_{ade}f_{ebc}(\delta_{ij}\delta_{kl}-\delta_{ik}\delta_{jl})$
\end{center}
\begin{center}
$+f_{ace}f_{ebd}(\delta_{ij}\delta_{kl}-\delta_{il}\delta_{jk}))$
\end{center}

\noindent\begin{picture}(100,50)(0,25)
\Gluon(30,25)(10,45){3}{4}
\Gluon(30,25)(10,5){3}{4}
\Line(30,25)(50,45)
\Line(30,25)(50,05)
\Vertex(10,5){3}
\Text(5,45)[r]{a,i}
\Text(5,5)[r]{b,j}
\Text(55,45)[l]{c}
\Text(55,5)[l]{d}
\end{picture}
$=-g^2(f_{ade}f_{ebc}+f_{ace}f_{ebd})\delta_{ij}$\\ \\ \\ \\

\noindent\begin{picture}(100,50)(0,25)
\Gluon(30,25)(10,45){3}{4}
\Gluon(30,25)(10,5){3}{4}
\Gluon(30,25)(50,45){3}{4}
\Gluon(30,25)(50,05){3}{4}
\Vertex(10,5){3}
\Vertex(10,45){3}
\Text(5,45)[r]{a,i}
\Text(5,5)[r]{b,j}
\Text(55,45)[l]{c,k}
\Text(55,5)[l]{d,l}
\end{picture}
$=-g^2(f_{abe}f_{ecd}(\delta_{ik}\delta_{jl}-\delta_{il}\delta_{jk}+\delta_{ij}{\delta_{kl}\over \xi})$

\begin{center}
$+f_{ade}f_{ebc}(\delta_{ij}\delta_{kl}-\delta_{ik}\delta_{jl}-\delta_{il}{\delta_{jk}\over
  \xi})$
\end{center}
\begin{center}
$+f_{ace}f_{ebd}(\delta_{ij}*\delta_{kl}-\delta_{il}\delta_{jk}))$
\end{center}

\noindent\begin{picture}(100,50)(0,25)
\Gluon(30,25)(10,45){3}{4}
\Gluon(30,25)(10,5){3}{4}
\Line(30,25)(50,45)
\Line(30,25)(50,05)
\Vertex(10,5){3}
\Vertex(10,45){3}
\Text(5,45)[r]{a,i}
\Text(5,5)[r]{b,j}
\Text(55,45)[l]{c}
\Text(55,5)[l]{d}
\end{picture}
$=-g^2(f_{ade}f_{ebc}+f_{ace}f_{ebd})\delta_{ij}$\\ \\ \\ \\

The two loop graphs to be computed are shown in Fig.~2.


The most naive way to perform the two-loop computation 
is by following the below vade-mecum.

\noindent First:
\begin{itemize}
\item contract the indices
\item scalarise integrals with no irreducible numerators using FORM~\cite{form}.
\end{itemize}

\noindent Then:

\begin{itemize}
\item Scalarise those which have irreducible numerators (to be defined later)
\end{itemize}

The whole result can be expressed in terms of 8 scalar integrals
. We have to compute them up to $\epsilon$ and expand 
in terms of $\tilde{p} = {p\over m_E}$. The result has the form:

\be 
\Pi_{ij} = \Pi(p^2,\xi,d,m^2)(p^2\delta_{ij} - p_i p_j) + C p_i p_j
\ee

\noindent where following our previous convention,

\be
lim_{\tilde p \to 0}\Pi(p^2,\xi,d,m^2)= F_2^{tr}
\ee

We have to check:

\begin{itemize}
\item the transversality (i.e. $C=0$)
\item gauge independence
\item U.V. convergence
\item I.R. convergence
\end{itemize}

At this point graphs 8,11,12,13,14 are zero in dimensional regularisation due to the presence
of a massless loop.
Diagrams 15 and 16 are one-loop diagrams due to the renormalization of the gauge parameter ~\cite{abbott}.

The main body of the problem is to express the whole two-loop graphs in term of scalar integrals
tabulated in the literature.

\section{Scalarisation, transversality and the two loop renormalisation}\label{sec:transv}

\begin{figure}

\begin{picture}(90,30)(0,0)
\SetScale{0.3}
\Gluon(110,50)(70,50){5}{4}
\Gluon(190,50)(230,50){5}{4}
\CArc(150,50)(40,90,270)
\CArc(150,50)(40,-90,90)
\Gluon(150,90)(150,10){5}{4}
\Vertex(70,50){4}
\Vertex(230,50){4}
\Text(45,-30)[]{Graph 1}
\end{picture}

\begin{picture}(90,30)(-90,-30)
\SetScale{0.3}
\Gluon(110,50)(70,50){5}{4}
\Gluon(190,50)(230,50){5}{4}
\Vertex(70,50){4}
\Vertex(230,50){4}
\GlueArc(110,10)(40,0,90){5}{4}
\CArc(150,50)(40,0,360)
\Text(45,-30)[]{Graph 2}
\end{picture}

\begin{picture}(90,30)(-180,-60)
\SetScale{0.3}
\Gluon(110,50)(70,50){5}{4}
\Gluon(190,50)(230,50){5}{4}
\Gluon(120,70)(180,70){3}{4}
\Vertex(70,50){4}
\Vertex(230,50){4}
\CArc(150,50)(40,0,360)
\Text(45,-30)[]{Graph 3}
\end{picture}

\begin{picture}(90,30)(-270,-90)
\SetScale{0.3}
\Gluon(110,50)(70,50){5}{4}
\Gluon(190,50)(230,50){5}{4}
\Vertex(70,50){4}
\Vertex(230,50){4}
\GlueArc(150,50)(40,-210,30){5}{10}
\CArc(150,50)(40,30,150)
\Line(120,70)(180,70)
\Text(45,-30)[]{Graph 4}
\end{picture}

\begin{picture}(90,30)(0,0)
\SetScale{0.3}
\Gluon(110,50)(70,50){5}{4}
\Gluon(190,50)(230,50){5}{4}
\GlueArc(150,50)(40,90,270){5}{10}
\CArc(150,50)(40,-90,90)
\Line(150,90)(150,10)
\Vertex(70,50){4}
\Vertex(230,50){4}
\Text(45,-30)[]{Graph 5}
\end{picture}

\begin{picture}(90,30)(-90,-30)
\SetScale{0.3}
\Gluon(150,10)(70,10){5}{8}
\Gluon(150,10)(230,10){5}{8}
\Gluon(120,70)(180,70){3}{4}
\Vertex(70,10){4}
\Vertex(230,10){4}
\CArc(150,50)(40,0,360)
\Text(45,-30)[]{Graph 6}
\end{picture}

\begin{picture}(90,30)(-180,-60)
\SetScale{0.3}
\Gluon(110,50)(70,50){5}{4}
\Gluon(190,50)(230,50){5}{4}
\Gluon(110,50)(190,50){5}{4}
\Vertex(70,50){4}
\Vertex(230,50){4}
\CArc(150,50)(40,0,360)
\Text(45,-30)[]{Graph 7}
\end{picture}

\begin{picture}(90,30)(-270,-90)
\SetScale{0.3}
\Gluon(150,10)(70,10){5}{8}
\Gluon(150,10)(230,10){5}{8}
\GlueArc(150,113)(20,0,360){5}{10}
\Vertex(70,10){4}
\Vertex(230,10){4}
\CArc(150,50)(40,0,360)
\Text(45,-30)[]{Graph 8}
\end{picture}

\begin{picture}(90,30)(0,0)
\SetScale{0.3}
\Gluon(150,10)(70,10){5}{8}
\Gluon(150,10)(230,10){5}{8}
\Vertex(70,10){4}
\Vertex(230,10){4}
\GlueArc(150,50)(40,-210,30){5}{10}
\CArc(150,50)(40,30,150)
\Line(120,70)(180,70)
\Text(45,-30)[]{Graph 9}
\end{picture}

\begin{picture}(90,30)(-90,-30)
\SetScale{0.3}
\Gluon(110,50)(70,50){5}{4}
\Gluon(190,50)(230,50){5}{4}
\GlueArc(150,50)(40,0,360){5}{10}
\Vertex(70,50){4}
\Vertex(230,50){4}
\CArc(150,113)(20,0,360)
\Text(45,-30)[]{Graph 10}
\end{picture}

\begin{picture}(90,30)(-180,-60)
\SetScale{0.3}
\Gluon(90,50)(70,50){5}{4}
\Gluon(210,50)(230,50){5}{4}
\GlueArc(120,50)(30,0,360){5}{10}
\CArc(180,50)(30,0,360)
\Vertex(70,50){4}
\Vertex(230,50){4}
\Text(45,-30)[]{Graph 11}
\end{picture}

\begin{picture}(90,30)(-270,-90)
\SetScale{0.3}
\Gluon(110,50)(70,50){5}{4}
\Gluon(190,50)(230,50){5}{4}
\Vertex(70,50){4}
\Vertex(230,50){4}
\CArc(110,10)(40,0,90)
\GlueArc(150,50)(40,-90,180){5}{10}
\CArc(150,50)(40,180,270)
\Text(45,-30)[]{Graph 12}
\end{picture}

\begin{picture}(90,30)(0,0)
\SetScale{0.3}
\Gluon(150,10)(70,10){5}{8}
\Gluon(150,10)(230,10){5}{8}
\CArc(150,113)(20,0,360)
\Vertex(70,10){4}
\Vertex(230,10){4}
\GlueArc(150,50)(40,0,360){5}{10}
\Text(45,-30)[]{Graph 13}
\end{picture}

\begin{picture}(90,30)(-90,-30)
\SetScale{0.3}
\Gluon(110,50)(70,50){5}{4}
\Gluon(190,50)(230,50){5}{4}
\CArc(150,50)(40,0,360)
\Vertex(70,50){4}
\Vertex(230,50){4}
\GlueArc(150,113)(20,0,360){5}{10}
\Text(45,-30)[]{Graph 14}
\end{picture}

\begin{picture}(90,30)(-180,-60)
\SetScale{0.3}
\Gluon(110,50)(70,50){5}{4}
\Gluon(190,50)(230,50){5}{4}
\GlueArc(150,50)(40,0,360){5}{10}
\Vertex(70,50){4}
\Vertex(230,50){4}
\BBoxc(110,50)(25,25)
\Text(45,-30)[]{Graph 15}
\end{picture}

\begin{picture}(90,30)(-270,-90)
\SetScale{0.3}
\Gluon(110,50)(70,50){5}{4}
\Gluon(190,50)(230,50){5}{4}
\GlueArc(150,50)(40,0,360){5}{10}
\Vertex(70,50){4}
\Vertex(230,50){4}
\BBoxc(150,90)(25,25)
\Text(45,-30)[]{Graph 16}
\end{picture}
\caption{ Two-loop Feynman graphs for the background two-point function. The boxes in $15$ and
  $16$ represent the one-loop renormalisation of the gauge parameter.} 
\end{figure}

\subsection{Scalarization}

The key point of the computation is to go from a set of integrals with momenta in the
numerator to a set of "master integrals" of a few scalar integrals ($8$ here) to compute. We are going
here to review the two methods of scalarisation that had been used in our two-loop
computation. These two methods are then implemented in the Mathematica routine TARCER~\cite{tarcer}. They
consist of kinematical scalarisation on one hand and in a identity based on integration by
parts on the other hand.

Indeed, as our vertices are vectorial, after their contraction with propagators, 
we obtain for each graph the integral over internal momenta of a product of propagators times a numerator.
The goal is then to cancel all the terms in the numerator with propagators.
At first sight, a natural way to do this is the so-called "kinematical scalarization".
It consists of writing the scalar products of momenta in the
numerator as a combination of propagators.\\
Let us take an example.\\
Let $q$ and $r$ be the internal momenta, $2q.r$ can be written as $ -((q-r)^2 + m^2) + (r^2 +m^2) + (q^2 +m^2) -  m^2$.
We see here that two situations appear. If all the terms in the rewriting of numerators cancel with the propagators,
a scalar integral will remain. This is what we call a reducible integral.
If it is not the case, i.e. if a term in the numerator is not present in the denominator, the method fails.
This is what we call a irreducible integral. For this latter, an other algorithm must be used
~\cite{Tarasov:1997kx}~\cite{Tarasov:1996br}.
Its non trivial part  relies on the following identity ~\cite{Tarasov:1997kx}~\cite{Tarasov:1996br}:

\begin{equation}
\prod_{i=1}^{L} \int d^dk_i
\frac{\partial }{\partial k_{ri}}
\left\{ R_{\{ i\} }\left( \{ k \}, \{ q \} \right)
\prod^{N}_{j=1} 
P^{\nu_j}_{\overline{k}_j,m_j} \right\} \equiv 0
\label{ibpm}
\end{equation}

\noindent where R is a polynomial function of internal and external momenta, L the number of
loops in the diagram and P are the various propagators power $\nu_j$ in the loop.

Those rules have been implemented in TARCER~\cite{tarcer}. We first use it in order to demonstrate
the transversality of our result. The integral to appear are ( with $d=3-2\epsilon$):

\TAIs $= \int {d^dq\over (2 \pi) ^3} {1 \over q^2 + m_E^2}$\\

\TBIms $= \int {d^dq\over (2 \pi) ^3}{1 \over q^2 + m_E^2}{1 \over (p+q)^2 + m_E^2}$\\

\TBIos $= \int {d^dq\over (2 \pi) ^3}{1 \over q^2 }{1 \over (p+q)^2}$\\

\TJIms $= \int \int {d^dq\over (2 \pi) ^3}{d^dr\over (2 \pi) ^3}{1 \over q^2 + m_E^2}{1 \over (p+r)^2 + m_E^2}{1 \over (q-r)^2 } $\\

\TJImos $=\int \int {d^dq\over (2 \pi) ^3}{d^dr\over (2 \pi) ^3}{1 \over q^2 + m_E^2}{1 \over r^2 + m_E^2}{1 \over (q-r)^2 }$\\
 
\TJIdms $=  \int \int {d^dq\over (2 \pi) ^3}{d^dr\over (2 \pi) ^3}{1 \over (q^2 + m_E^2)^2}{1 \over r^2 + m_E^2}{1 \over (q-r)^2 }$\\

\TVIs $ =   \int \int {d^dq\over (2 \pi) ^3}{d^dr\over (2 \pi) ^3} {1 \over ((q+p)^2 + m_E^2)^2}{1 \over r^2 + m_E^2}{1 \over (q-r)^2 }$\\

\TFIs $ =   \int \int {d^dq\over (2 \pi) ^3}{d^dr\over (2 \pi) ^3}{1 \over q^2 }{1 \over r^2 +
  m_E^2}{1 \over (q-r)^2 + m_E^2 }{1 \over (q+p)^2}{1 \over (r+p)^2 + m_E^2}$\\
\\
Their values are given in Appendix 1.

\subsection{Transversality for any dimension d}\label{sec:trans}

The goal of this section is to check that our result is transverse.
This is due to the fact that we compute a self energy using the background
method. Before presenting the result, it must be noticed that only the scalarisation process
is necessary to prove this. Indeed , one need not to express the master integrals in terms of
external momenta, even not to specify the dimension of space. So our result is general.
Let us write

\be
\tilde{C} = p_i \Pi_{ij} p_j = p^4 C
\ee

\noindent and check that C is zero.

After scalarisation, all the non transverse component can be expressed in terms 
of three master integrals. These master integrals are tabulated in Appendix 1.  
As defined in the last subsection, the propagators and vertices are all scalars
. Thin lines represent massless propagators and bold line represent massive propagators of
mass $m_E$. For each diagram, the result is given in Appendix 2 in terms of the master integrals.
Here is the result for the first graph.

\bea
\nnb&\tilde{C}_1=&\\
\nnb&-&{d-d^2+6\xi-5d\xi+d^2\xi\over12(-3+d)}{(\TAIcarre)^2}+\\
\nnb&-&{-24m^2+34dm^2-10d^2m^2-2p^2+3dp^2\over12(-2+d)}{\TJIm}\\
\nnb&-&{-d^2p^260m^2\xi-50dm^2\xi+10d^2m^2\xi\over12(-2+d)}{\TJIm}\\
\nnb&-&{10p^2\xi-7dp^2\xi+d^2p^2\xi\over12(-2+d)}{\TJIm}\\ 
\nnb&-&{m^2(4m^2+p^2)(1-d-2\xi+d\xi)\over3(-2+d)}{\TJIm}\\
\eea

What we have learnt is that, whereas the whole result is transverse, 
individuals graphs are not. Moreover, the process of scalarization is sufficient 
to show explicitely transversality. It will not be the case to show gauge 
independance of the coupling, where the expressions of master integrals are needed. 

When all non transverse contributions are summed up, then the whole result is $0$ (see Appendix
2). The latter of course is true for any dimension.

\subsection{Scalarisation of the transverse part}\label{sec:trans2}

Now that we have demonstrated that our result is transverse, we want to compute
the renormalisation of the coupling. We have to scalarize the transverse part. The complexity
of the following computation is due to the presence of a general gauge parameter $\xi$ in the
Lagrangian, that is reflected by the gauge dependence of the magnetic gluon propagator. But also,
due to the presence of a magnetic background, to the gauge dependence of the vertices.
Let us call $Fi_2$ the value of a two-loop graph contracted with $\delta_{ij}$. 
We have
\be
\sum_i Fi_2 = p^2 F_2^{tr}
\ee

For the first graph, the result reads:

\bea
\nnb&F1_2=&\\
\nnb&-&{48m^4-96dm^4+60d^2m^4\over12(-4+d)(-3+d)m^2p^2(4m^2+p^2)}{(\TAIcarre)^2}\\
\nnb&-&{-12d^3m^4+36m^2p^2-50dm^2p^2\over12(-4+d)(-3+d)m^2p^2(4m^2+p^2)}{(\TAIcarre)^2}\\
\nnb&-&{17d^2m^2p^2-3d^3m^2p^2+12p^4-6dp^4\over12(-4+d)(-3+d)m^2p^2(4m^2+p^2)}{(\TAIcarre)^2}\\
\nnb&-&{-144m^4\xi+192dm^4\xi-84d^2m^4\xi\over12(-4+d)(-3+d)m^2p^2(4m^2+p^2)}{(\TAIcarre)^2}\\
\nnb&-&{12d^3m^4\xi-36m^2p^2\xi+48dm^2p^2\xi\over12(-4+d)(-3+d)m^2p^2(4m^2+p^2)}{(\TAIcarre)^2}\\
\nnb&-&{-21d^2m^2p^2\xi+3d^3m^2p^2\xi\over12(-4+d)(-3+d)m^2p^2(4m^2+p^2)}{(\TAIcarre)^2}\\
\nnb&-&{4m^2-2p^2+dp^2\over2(-4+d)(-3+d)m^2}{\TAI}{\TBIm}\\
\nnb&-&{2m^2+p^2\over2(-4+d)}{(\TBImcarre)^2}\\
\nnb&-&{384m^4-720dm^4+408d^2m^4-72d^3m^4\over12(-4+d)(-2+d)p^2(4m^2+p^2)}{\TJIm}\\
\nnb&-&{144m^2p^2-200dm^2p^2+134d^2m^2p^2\over12(-4+d)(-2+d)p^2(4m^2+p^2)}{\TJIm}\\
\nnb&-&{-30d^3m^2p^2-32p^4+18dp^4+5d^2p^4-3d^3p^4\over12(-4+d)(-2+d)p^2(4m^2+p^2)}{\TJIm}\\
\nnb&-&{-1152m^4\xi+1392dm^4\xi-552d^2m^4\xi+72d^3m^4\xi\over12(-4+d)(-2+d)p^2(4m^2+p^2)}{\TJIm}\\
\nnb&-&{-480m^2p^2\xi+564dm^2p^2\xi-222d^2m^2p^2\xi\over12(-4+d)(-2+d)p^2(4m^2+p^2)}{\TJIm}\\
\nnb&-&{30d^3m^2p^2\xi-48p^4\xi+54dp^4\xi-21d^2p^4\xi\over12(-4+d)(-2+d)p^2(4m^2+p^2)}{\TJIm}\\
\nnb&-&{3d^3p^4\xi\over12(-4+d)(-2+d)p^2(4m^2+p^2)}{\TJIm}\\
\nnb&-&{36m^4-66dm^4+36d^2m^4-6d^3m^4-11dm^2p^2\over3(-4+d)(-3+d)(-2+d)p^2}{\TJIdm}\\
\nnb&-&{14d^2m^2p^2-3d^3m^2p^2-6p^4+3dp^4\over3(-4+d)(-3+d)(-2+d)p^2}{\TJIdm}\\
\nnb&-&{-108m^4\xi+126dm^4\xi-48d^2m^4\xi+6d^3m^4\xi\over3(-4+d)(-3+d)(-2+d)p^2}{\TJIdm}\\
\nnb&-&{-36m^2p^2\xi+48dm^2p^2\xi-21d^2m^2p^2\xi+3d^3m^2p^2\xi\over3(-4+d)(-3+d)(-2+d)p^2}{\TJIdm}\\
\nnb&+&{(-1+d)p^2(2m^2+3p^2)\over12(-1+d)m^2p^2(4m^2+p^2)}{(\TAIcarre)^2}\\
\nnb&-&{(-1+d)p^2\over4(-1+d)m^2}{\TAI}{\TBIm}\\
\nnb&+&{(-1+d)p^2\over2(-1+d)(4m^2+p^2)}{\TJImo}\\
\nnb&-&{(-1+d)p^2(-12m^4+4dm^4-14m^2p^2+4dm^2p^2)\over12(-2+d)(-1+d)m^2p^2(4m^2+p^2)}{\TJIm}\\
\nnb&-&{(-1+d)p^2(2m^2+3p^2)\over6(-2+d)(-1+d)p^2}{\TJIdm}\\
\nnb&-&{(-1+d)p^2\over2(-1+d)}{\TVI}\\
\eea

Considering the example above, further general remarks can be made.
First, compared with the non-transverse part of the first graph that has been shown previously,
the transverse part involves more master integrals and is a little bit more
complicated. Moreover, in the scalarisation process, some poles in $\epsilon$ appear.
They will of course cancel, witnessing U.V. convergence. But attention has to be paid
when one has to compute the final result. Indeed some final parts when $\epsilon \to 0$ are
provided by the product of a $1 \over \epsilon$ coming from scalarization and the first term in
$\epsilon$ of a master integral.
Indeed not all the master integrals have to be computed up and includind $O(\epsilon)$. 

\subsection{Result}

Once one has computed the master integrals and develop them for $p^2 \ll m_E^2$, the table in
Fig.~3 is obtained. 
\begin{figure}
\begin{tabular}{|c|c|}
\hline
\raisebox{0pt}[14pt][6pt]{Graph no} &
\raisebox{0pt}[14pt][6pt]{$p^2 F_2^{tr}$} \\
\hline 
&\\
1& $-{\tilde{p}^2(29+24\xi)\over9216\pi^2}+{1+\xi+\epsilon(2+\xi-4(1+\xi)\log(2))+4\epsilon(1+\xi)\log({\bar{\mu}\over m})\over128\epsilon\pi^2}$ \\
2& ${\tilde{p}^2(20+6\xi)\over1536\pi^2}-{3(2+\xi-2\epsilon(-1+\xi(-1+\log(4))+\log(16))+4\epsilon(2+\xi)\log({\bar{\mu}\over m})))\over128\epsilon\pi^2}$ \\
3& ${\tilde{p}^2(-43+12\xi)\over4608\pi^2}+{4+\epsilon(1+\xi-16\log(2))+16\epsilon\log({\bar{\mu}\over m})\over64\epsilon\pi^2}$ \\
4& ${-2-\xi\over32\tilde{p}\pi}-{\tilde{p}(15+\xi))\over1536\pi}+{\tilde{p}^2(13+2\xi)\over3072\pi^2}-{3(1+\xi)^2+4\epsilon(-8+\xi)\xi\log({\bar{\mu}\over2m})+4\epsilon(3+2\xi
(7+\xi))\log({\bar{\mu}\over2m})\over768\epsilon\pi^2\xi}$ \\
5& $ {\tilde{p}(15+\xi(4+\xi))\over1536\pi}-{\tilde{p}^2(23+12\xi)\over4608\pi^2}+{2\epsilon+\xi+4\epsilon\xi\log({\bar{\mu}\over2m})\over128\epsilon\pi^2}$ \\
6& ${-3\over(64\epsilon\pi^2)}+{22-24(1+\log({\bar{\mu}\over2m}))\over128\pi^2}$ \\
7& ${6(2+\xi)+\epsilon(12(1+\xi)-\tilde{p}^2(2+\xi))+24\epsilon(2+\xi)
\log({\bar{\mu}\over2m})\over512\epsilon\pi^2}$\\
8& $0$\\
9& ${1+2\xi\over256\epsilon\pi^2\xi}+{-2\xi-4(1+2\xi)-4(-1-2\xi)(1+\log({\bar{\mu}\over2m}))\over256\pi^2\xi}$\\
10& ${16+8\xi\over256\tilde{p}\pi}$\\
11& $0$\\
12& $0$\\
13& $0$\\
14& $0$\\
15& $-{\tilde{p}(1+\xi)(4+\xi)\over1536\pi}$\\
16& ${\tilde{p}(2+\xi)\over768\pi}$\\
&\\
\hline
\hline
&\\
sum& ${-19\tilde{p}^2\over4608\pi^2}$\\
&\\
\hline
\end{tabular}
\caption{ Final result for each graph up to $O(\tilde p^2)$.}  
\end{figure}
So finally 
\be 
\sum_i Fi_2 = p^2 F_2^{tr} = {-19\tilde{p}^2\over4608\pi^2}
\ee

This result has of course no pole in $\epsilon$, is $\xi$ independant, has no pole in
$\tilde{p}$, is $\mu$ independant whereas each graph does not have these properties. This is a really strong check
of our computation. Going into more detail  will show that this check is robust in the sense
that an error in a integral will show up through one of this checks. So the result is simply:

\be
g_M^{(2)} = {-19\over4608\pi^2m_{E}^2}g_{E}^4 N^2
\ee

\noindent Our main result, eq.(\ref{final}), shows that the smallness of the corrections to the magnetic coupling does persist in two-loop order. In fact, at $2T_c$ the coupling for 3 colours equals~\cite{owe}  $g^2_E=2.7$ and the two-loop correction is about a third of the one loop correction (itself about 3 percent).

Our result is of importance in analyzing the  purely magnetic quantities,
such as the spatial Wilson loop, and the magnetic mass. In particular it is crucial in connecting the lattice results from the magnetic action  to those obtained 
from the electric action, and ultimately to those of four dimensional simulations.

\section{Spatial Wilson Loop}

The goal of this part is to test the applicability of 3D physics at intermediate T.

As has been recalled previously, the non-perturbative part in the weak
coupling expansion of an observable comes from its magnetostatic thermal average.
The dominant term of this non-perturbative part comes from the $3D$ dimensional 
Yang-Mills with its coupling computed a zero-loop order.
Its next term is a $3D$ dimensional 
Yang-Mills with its coupling computed a one-loop order.
Then a three-dimensional 
Yang-Mills with its coupling computed a two-loop order. Then the next term is the mix of 
a three-dimensional 
Yang-Mills with its coupling computed at three-loop order and higher order operators.

Our actual knowledge of the non-perturbative part is now the 3D Yang-Mills action with its coupling
computed at two-loop order. The natural question is: does this precision encompass almost 
the whole non perturbative contribution to the thermal average of an observable ?

Let us consider an observable which has its dominant contribution from the 3D Yang-Mills
sector, unlike the pressure whose dominant contribution is the 
Stefan-Boltzmann term due to hard modes. That means that at infinite temperature, its thermal average is computed with
the zero-loop magnetostatic Lagrangian, namely the three-dimensional Yang-Mills action.

Such an observable is the spatial Wilson loop in the fundamental representation:
\begin{equation}
W(L)=Tr{P}\exp{(i\oint_L g\vec A \cdot d\vec l)}.
\label{wilsonloop}
\end{equation}

\noindent As $L$ is purely spatial, it measures the magnetic flux in the plasma.  The thermal average of this spatial Wilson loop shows area behaviour with a surface tension $\sigma(T)$. As this is a purely magnetic quantity, we expect from dimensional arguments that $\sqrt{\sigma} = c g_M^2$ where $c$ is a nonperturbative proportionality constant.
Indeed, as the average of the loop is due to long distance correlations, hard modes
will not have any effect on the thermal average. In the same way,
for soft modes, we can integrate out the $A_0$ field,
which is what we have done in the previous section while constructing $L_M$,
as the loop does not depend on $A_0$ ! Finally, we have:

\begin{equation}
\langle W(L) \rangle = \exp{-\sigma A(L)}=\frac{\int  D\vec A W(L)\exp{-S_M(A)}}{\int D\vec A\exp{-S_M(A)}},
\label{eq:wloopmagnetic}
\end{equation}

\noindent and it gives, as $\delta L_M$ is of relative order $g^3$,
\begin{equation}
\sigma(T)=c^2g_M^4(1+O(g^3)).
\label{eq:wtensionmagnetic}
\end{equation}

Now the aim is to fit with our formula for $g_M$ the proportionality constant $c$. For that, we have to go to the lattice. We are going to fit
$ T\over \sqrt{\sigma}$ as a function of $T \over T_c $. Indeed, $g_M$ is a function of $g_E$ and $m_E$ which
are functions of $g$ and $T$.
So, for $N = 3$:

\begin{equation}
g_M^{-2} = g^{-2} T^ {-1} (1 + {g\over 16 \pi} + {19 g^2 \over 512 \pi^2})
\end{equation}

\noindent with

\begin{equation}
g^{-2} = {11 \over 8 \pi^2}(Log( {T \over T_c}) +  Log( {T_c \over \Lambda_{\overline{MS}}}) + 1.90835... ).
\end{equation}

\begin{figure}[t]
\begin{center}
\includegraphics{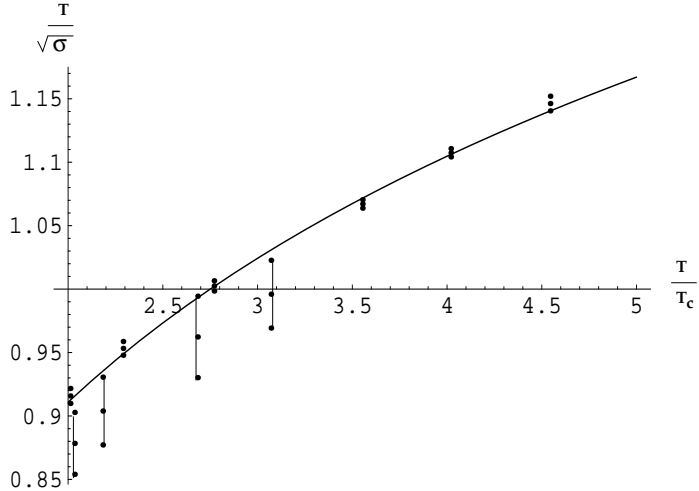}
\caption{Plot of  $T\over \sqrt{\sigma}$  versus $T \over T_c$ and fit with our two-loop formula. Each of the 10 points is shown with its error bar.}
\label{fig:fit1}
\end{center}
\end{figure}

\noindent The $\mu$ dependance will be discussed in next section. 
So thanks to these formulas, one can fit $ T\over \sqrt{\sigma}$  versus $T \over T_c$
in order to determin  $c$ and $T_c \over \Lambda_{\overline{MS}}$. Data points for the $SU(3)$
spatial string tension has been taken from ~\cite{Boyd:1996bx,Lutgemeier:xi} for $T > 2 T_c$. So we have $10$ points. The fit is shown in fig~\ref{fig:fit1}.

The fit gives:

\begin{eqnarray}
{T_c\over{\Lambda_{\overline{MS}}}} &=& 1.78(12)\\
c &=& 0.505(11).
\end{eqnarray}

\noindent In order to estimate the error in our fit, we have plotted for those $10$ points the
confidence levels taken from the $\chi^2$ distribution. 
Compared with the results in the literature our values are incompatible with them.

\begin{figure}[t]
\begin{center}

\includegraphics{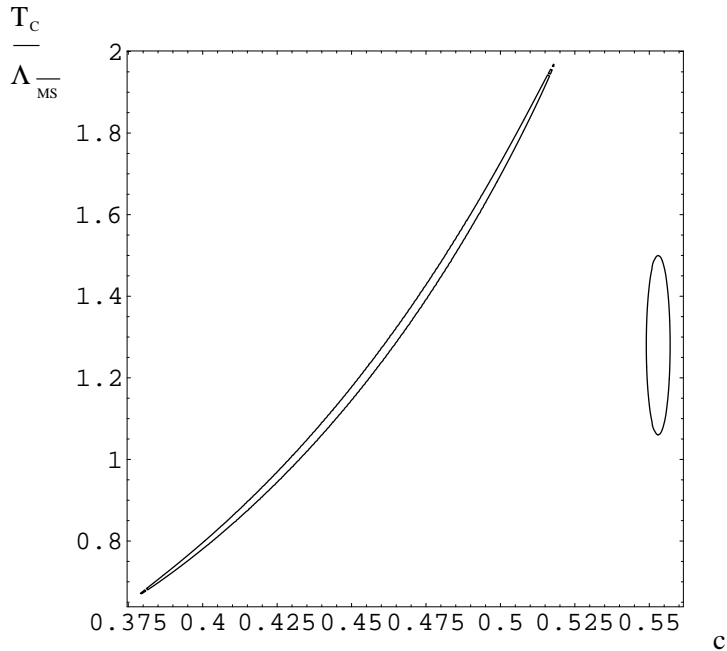}

\caption{ Plot of various confidence levels in the plane c- $T_c \over \Lambda_{\overline MS}$
  for the $4$ data points with the smallest error bar.
The widespread ellipse on the left delimits the $95$ percent confidence level for the parameters.
The central values are $c = 0.481$ and  ${T_c \over \Lambda_{\overline{MS}}} = 1.27$.
The ellipse on the r.h.s. represents the actual values for these parameters from the
literature ~\protect\cite{necco,Teper:1998te,Lucini:2002wg}.}
\label{fig:quifit}
\end{center}
\end{figure}

Then we have also fitted data with $T > 2.5 T_c$. This leaves us with $6$ points.
We obtained
\begin{eqnarray}
{T_c\over{\Lambda_{\overline{MS}}}} &=& 1.57(20)\\
c  &=& 0.488(18)
\end{eqnarray}

\noindent We have plotted for the $4$ more precise of those $6$ points the $95$ percent confidence level contour taken
from the $\chi^2$ distribution in Fig.~5.
Let us draw our conclusions.

Despite the fact that the region of confidence is larger, the value of 
${T_c\over{\Lambda_{\overline{MS}}}}$ and 
c is still incompatible with refs.~\cite{Teper:1998te,Lucini:2002wg,Karsch:1994af}.
To see compatibility one obviously needs string data for higher $T$. It might also be recommendable to use other Monte Carlo updates, like the one described in ref.~\cite{HMeyer}, which is a version of the L\"uscher-Weisz algorithm~\cite{weisz}.

\subsection{Renormalisation scale dependence}

In the previous section, we have taken implicitely, as renormalisation scale,
the one that cancels the one loop renormalisation due to hard modes of
the electric coupling, namely $\tilde{\mu_T}= \mu_T = 4 \pi T e^{- \gamma_E} e^{{-1\over 11}}$~\cite{huang}.

This $\mu$ dependence appears explicitely when one tries to express the magnetic coupling $g_M$
as a function of the QCD coupling $g$. In fact the renormalization scale will appear 
in two ways. The first is through the renormalization of the QCD coupling, and the second is in
the expression of $g_E$ as a function of $g$.

Namely, for a pure gauge theory, the electrostatic coupling reads

\be
g_E^2 =  g^2(\mu)T(1+{11N\over24}(Log({\mu\over \tilde \mu_T}))
\ee

\noindent where

\be
g^2(\mu)= {24 \pi^2 \over 11 N ln (\mu / \Lambda_{\overline{MS}})}
\ee

\noindent If one combines those two equations, evaluating in $\mu = \tilde \mu_T + \delta \mu$, one finds that, in
our case with no fermions, the first term in $\delta \mu$ will disappear when one expand $g_E^2$
near $\mu =\tilde \mu_T$. 

\subsection{Conclusion}

We have seen that we have not yet captured the full $4D$ physics with our truncation. That means that
higher order terms are needed to make the link. This has been done recently by considering 
the two loop renormalization of $g_E$~\cite{mikkoyork}. The conclusion, also with the fact that we need
more lattice data,  is that at the temperatures that we have
considered, perturbative and non-perturbative effects are both important. The next step is to compute the magnetostatic Lagrangian at higher order, determining
for example the leading higher dimensional operators. This will be the purpose of a forthcoming paper~\cite{moibientot}

\section*{Acknowledgment}
I thank Chris Korthals Altes for his help throughout my work.  
I acknowledge the help of Mikko Laine and York Schr\"oder for very useful and stimulating
advice. I thank the DAAD and the University of Bielefeld for financial support.
\newpage

\section*{Appendix 1: Scalar Integrals}

In this appendix, we compute the master integrals that appear in the non-transverse part.
The computation is carried out for $d= 3- 2 \epsilon$. For $\epsilon = 0$, these integrals have been 
tabulated in Ref.~\cite{rajantie}. For some of them, when a ${1\over \epsilon}$ appears in the
process of scalarisation for one of the Feynman graphs, we have to compute the first term
in the $\epsilon$ expansion. This has been done using the same techniques as in
~\cite{rajantie}. At the end, one has not to forget that we are only interested in 
the result for $p^2 \ll m_E^2$ where $p$ is the external momentum.
We have used the same notations as in Ref.~\cite{rajantie}, namely, the measure of
integration is $\int{d\vec l\over{(2\pi)^d}}$ and thin lines represent massless propagators and bold line represent massive propagators of
mass $m_E$.\\

\TAIs $= - {m_E \over 4 \pi} (1 + 2\epsilon)(1 + \log{\bar \mu \over2m_E})) + O(\epsilon ^2)$\\

\TBIms $= {ArcTan{p\over 2m_E}\over 4p\pi} - (2\epsilon)R_1(p^2) + O(\epsilon ^2)$\\

\TBIos $= {1\over 8p} + O(\epsilon)$\\

\TJIms $= {1\over (4\pi)^2} ({1\over 4 \epsilon} + 3/2 - 2 m_E {ArcTan{p\over 2m_E} \over p}$\\
\begin{center}
$ + {Log{\bar{\mu}^2 \over p^2 + 4 m_E^2}\over 2}) + O(\epsilon)$\\
\end{center}

\TJImos $= 
  {1\over (4\pi)^2}({1 \over 4\epsilon} + 1/2 + Log{\bar{\mu} \over 2m_E}) + O(\epsilon) $\\

\TJIdms $= 
 {ArcTan{p\over 2m_E}\over(4\pi)^2} {1\over 2m_Ep} + (-2\epsilon)R_2(p^2)+ O(\epsilon ^2) $\\

For the two following master integrals, only the part in $\epsilon = 0$ is required. Moreover,
they have been developed in $\tilde{p} = {p \over m_E}$.

\TVIs $={1\over(4 m^2)} {1\over (4 \pi)^2} + O(p^2)$ \\

\TFIs $= {1 \over 256 \pi  p^2 m_E ^2} - {1\over 256  \pi^2  m_E^4 } + O(p^2) $\\

Then, the computation contained in the main body of this paper (see Appendix 3) shows that we have to determine
the first term in $\epsilon$ contained in two scalar integrals. Namely:

\be 
R_1(p^2)= {-1\over 2} {Log{\bar{\mu}^2 \over(2m_E)^2} \over 8\pi m_E } + 
        p^2 {-Log{\bar{\mu}^2 \over (2m_E)^2} - 2 \over 96 m_E^3 \pi} + O(p^4)
\ee

\be
R_2(p^2) = {1+Log{\bar{\mu}^2 \over (2m_E)^2} \over 32 m_E^2 \pi^2} +p^2 {-17 - 12  Log{\bar{\mu}\over (2m_E)}\over 4608 m_E^4 \pi^2} + O(p^4)
\ee

\newpage

\section*{Appendix 2}

In this Appendix, the non-transverse part, contracted with $p_i p_j$ is computed for each graph.\\

\noindent

$\tilde{C}_1=$
\bea
\nnb&-&{d-d^2+6\xi-5d\xi+d^2\xi\over12(-3+d)}{(\TAIcarre)^2}+\\
\nnb&-&{-24m^2+34dm^2-10d^2m^2-2p^2+3dp^2\over12(-2+d)}{\TJIm}\\
\nnb&-&{-d^2p^260m^2\xi-50dm^2\xi+10d^2m^2\xi\over12(-2+d)}{\TJIm}\\
\nnb&-&{10p^2\xi-7dp^2\xi+d^2p^2\xi\over12(-2+d)}{\TJIm}\\ 
\nnb&-&{m^2(4m^2+p^2)(1-d-2\xi+d\xi)\over3(-2+d)}{\TJIm}\\
\eea

$\tilde{C}_2=$
\bea
\nnb&-&{(-1+d)(-1+\xi)\over4}{(\TAIcarre)^2}\\
\nnb&+&{-3m^2+4dm^2-d^2m^2-2p^2+3dp^2-d^2p^2+3m^2\xi\over2(-2+d)}{\TJIm}\\
\nnb&+&{-4dm^2\xi+d^2m^2\xi+8p^2\xi-6dp^2\xi+d^2p^2\xi\over2(-2+d)}{\TJIm}\\
\nnb&+&{(-1+d)m^2(4m^2+p^2)(-1+\xi)\over2(-2+d)}{\TJIdm}\\
\eea

$\tilde{C}_3=$
\bea
\nnb&+&{-6m^2+10dm^2-4d^2m^2-6p^2+9dp^2\over12(-3+d)m^2}{(\TAIcarre)^2}\\
\nnb&+&{-3d^2p^2+18m^2\xi-18dm^2\xi+4d^2m^2\xi\over12(-3+d)m^2}{(\TAIcarre)^2}\\
\nnb&+&{(-18m^2+8dm^2+2p^2-dp^2)(1-d-3\xi+d\xi)\over6(-2+d)}{\TJIm}\\
\nnb&+&{m^2(4m^2+p^2)(1-d-3\xi+d\xi)\over3(-2+d)}{\TJIdm}\\
\eea

$\tilde{C}_4=$
\bea
\nnb&+&{-6p^2+9dp^2-3d^2p^2+15dm^2\xi-33d^2m^2\xi\over24(-5+d)(-3+d)dm^2\xi}{(\TAIcarre)^2}\\
\nnb&+&{21d^3m^2\xi-3d^4m^2\xi+6p^2\xi-15dp^2\xi\over24(-5+d)(-3+d)dm^2\xi}{(\TAIcarre)^2}\\
\nnb&+&{12d^2p^2\xi-3d^3p^2\xi-75dm^2\xi^2\over24(-5+d)(-3+d)dm^2\xi}{(\TAIcarre)^2}\\
\nnb&+&{85d^2m^2\xi^2-29d^3m^2\xi^2+3d^4m^2\xi^2\over24(-5+d)(-3+d)dm^2\xi)}{\TAIcarre)^2}\\ 
\nnb&+&{-21m^2+30dm^2-9d^2m^2+57m^2\xi-46dm^2\xi\over12(-2+d)}{\TJIm}\\
\nnb&+&{9d^2m^2\xi+2p^2\xi-dp^2\xi\over12(-2 + d)}{\TJIm}\\
\nnb&+&{m^2(4m^2+p^2)(3-3d-7\xi+3d\xi)\over12(-2+d)}{\TJIdm}\\
\eea

$\tilde{C}_5=$
\bea
\nnb&-&{-3+6d-3d^2+15\xi-14d\xi+3d^2\xi\over12(-3+d)}{(\TAIcarre)^2}\\
\nnb&-&{-21m^2+30dm^2-9d^2m^2+57m^2\xi\over6(-2+d)}{\TJIm}\\
\nnb&-&{-46dm^2\xi+9d^2m^2\xi+2p^2\xi-dp^2\xi\over6(-2+d)}{\TJIm}\\
\nnb&-&{m^2(4m^2+p^2)(3-3d-7\xi+3d\xi)\over6(-2+d)}{\TJIdm}\\
\eea

$\tilde{C}_6=$
\bea
\nnb&+&{(-2+d)(-1+d)p^2\over4(-3+d)m^2}{(\TAIcarre)^2}\\
\eea

$\tilde{C}_7=$
\bea
\nnb&+&{(-1+d)(-1+\xi)\over8}{(\TAIcarre)^2}\\ 
\nnb&-&{-3m^2+4dm^2-d^2m^2-2p^2\over4(-2+d)}{\TJIm}\\
\nnb&-&{3dp^2-d^2p^2+3m^2\xi-4dm^2\xi+d^2m^2\xi\over4(-2+d)}{\TJIm}\\
\nnb&-&{8p^2\xi-6dp^2\xi+d^2p^2\xi\over4(-2+d)}{\TJIm}\\ 
\nnb&-&{(-1+d)m^2(4m^2+p^2)(-1+\xi)\over(4(-2+d)}{\TJIdm}\\
\eea

$\tilde{C}_9=$
\be
{(-2+d)(-1+d)p^2(1-\xi+d\xi)\over8(-5+d)(-3+d)dm^2\xi}{(\TAIcarre)^2}\\
\ee

\noindent$\tilde{C}_{10} = 0$\\
$\tilde{C}_{15}= 0$\\ 
$\tilde{C}_{16}=0$\\ 

When all the contributions are summed up, the result is zero.

\newpage
\section*{Appendix 3}

In this section the expression of the transverse part of each graph is given in terms of 
the 8 scalar integrals: 

$F1_2=$
\bea
\nnb&-&{48m^4-96dm^4+60d^2m^4\over12(-4+d)(-3+d)m^2p^2(4m^2+p^2)}{(\TAIcarre)^2}\\
\nnb&-&{-12d^3m^4+36m^2p^2-50dm^2p^2\over12(-4+d)(-3+d)m^2p^2(4m^2+p^2)}{(\TAIcarre)^2}\\
\nnb&-&{17d^2m^2p^2-3d^3m^2p^2+12p^4-6dp^4\over12(-4+d)(-3+d)m^2p^2(4m^2+p^2)}{(\TAIcarre)^2}\\
\nnb&-&{-144m^4\xi+192dm^4\xi-84d^2m^4\xi\over12(-4+d)(-3+d)m^2p^2(4m^2+p^2)}{(\TAIcarre)^2}\\
\nnb&-&{12d^3m^4\xi-36m^2p^2\xi+48dm^2p^2\xi\over12(-4+d)(-3+d)m^2p^2(4m^2+p^2)}{(\TAIcarre)^2}\\
\nnb&-&{-21d^2m^2p^2\xi+3d^3m^2p^2\xi\over12(-4+d)(-3+d)m^2p^2(4m^2+p^2)}{(\TAIcarre)^2}\\
\nnb&-&{4m^2-2p^2+dp^2\over2(-4+d)(-3+d)m^2}{\TAI}{\TBIm}\\
\nnb&-&{2m^2+p^2\over2(-4+d)}{(\TBImcarre)^2}\\
\nnb&-&{384m^4-720dm^4+408d^2m^4-72d^3m^4\over12(-4+d)(-2+d)p^2(4m^2+p^2)}{\TJIm}\\
\nnb&-&{144m^2p^2-200dm^2p^2+134d^2m^2p^2\over12(-4+d)(-2+d)p^2(4m^2+p^2)}{\TJIm}\\
\nnb&-&{-30d^3m^2p^2-32p^4+18dp^4+5d^2p^4-3d^3p^4\over12(-4+d)(-2+d)p^2(4m^2+p^2)}{\TJIm}\\
\nnb&-&{-1152m^4\xi+1392dm^4\xi-552d^2m^4\xi+72d^3m^4\xi\over12(-4+d)(-2+d)p^2(4m^2+p^2)}{\TJIm}\\
\nnb&-&{-480m^2p^2\xi+564dm^2p^2\xi-222d^2m^2p^2\xi\over12(-4+d)(-2+d)p^2(4m^2+p^2)}{\TJIm}\\
\nnb&-&{30d^3m^2p^2\xi-48p^4\xi+54dp^4\xi-21d^2p^4\xi\over12(-4+d)(-2+d)p^2(4m^2+p^2)}{\TJIm}\\
\nnb&-&{3d^3p^4\xi\over12(-4+d)(-2+d)p^2(4m^2+p^2)}{\TJIm}\\
\nnb&-&{36m^4-66dm^4+36d^2m^4-6d^3m^4-11dm^2p^2\over3(-4+d)(-3+d)(-2+d)p^2}{\TJIdm}\\
\nnb&-&{14d^2m^2p^2-3d^3m^2p^2-6p^4+3dp^4\over3(-4+d)(-3+d)(-2+d)p^2}{\TJIdm}\\
\nnb&-&{-108m^4\xi+126dm^4\xi-48d^2m^4\xi+6d^3m^4\xi\over3(-4+d)(-3+d)(-2+d)p^2}{\TJIdm}\\
\nnb&-&{-36m^2p^2\xi+48dm^2p^2\xi-21d^2m^2p^2\xi+3d^3m^2p^2\xi\over3(-4+d)(-3+d)(-2+d)p^2}{\TJIdm}\\
\nnb&+&{(-1+d)p^2(2m^2+3p^2)\over12(-1+d)m^2p^2(4m^2+p^2)}{(\TAIcarre)^2}\\
\nnb&-&{(-1+d)p^2\over4(-1+d)m^2}{\TAI}{\TBIm}\\
\nnb&+&{(-1+d)p^2\over2(-1+d)(4m^2+p^2)}{\TJImo}\\
\nnb&-&{(-1+d)p^2(-12m^4+4dm^4-14m^2p^2+4dm^2p^2)\over12(-2+d)(-1+d)m^2p^2(4m^2+p^2)}{\TJIm}\\
\nnb&-&{(-1+d)p^2(2m^2+3p^2)\over6(-2+d)(-1+d)p^2}{\TJIdm}\\
\nnb&-&{(-1+d)p^2\over2(-1+d)}{\TVI}\\
\eea

$F2_2=$
\bea
\nnb&+&{-18m^2+14dm^2-6p^2+3dp^2\over8(-3+d)m^2(4m^2+p^2)}{(\TAIcarre)^2}\\
\nnb&-&{3(-4m^2+4dm^2-2p^2+dp^2)\over8(-3+d)m^2}{\TAI}{\TBIm}\\
\nnb&-&{-18m^2+14dm^2-10p^2+5dp^2-48m^2\xi+24dm^2\xi-12p^2\xi+6dp^2\xi\over4(-2+d)(4m^2+p^2)}{\TJIm}\\
\nnb&-&{-18m^2+14dm^2-6p^2+3dp^2\over4(-3+d)(-2+d)}{\TJIdm}\\
\nnb&+&{3(-1+d)p^2(2m^2+3p^2)\over12(-1+d)m^2p^2(4m^2+p^2)2}{(\TAIcarre)^2}\\
\nnb&-&{3(-1+d)p^2\over4(-1+d)2m^2}{\TAI}{\TBIm}\\
\nnb&+&{3(-1+d)p^2\over2(-1+d)2(4m^2+p^2)}{\TJImo}\\
\nnb&-&{3(-1+d)p^2(-12m^4+4dm^4-14m^2p^2+4dm^2p^2)\over12(-2+d)2(-1+d)m^2p^2(4m^2+p^2)}{\TJIm}\\
\nnb&-&{3(-1+d)p^2(2m^2+3p^2)\over6(-2+d)2(-1+d)p^2}{\TJIdm}\\
\nnb&-&{3(-1+d)p^2\over2(-1+d)2}{\TVI}\\
\eea

$F3_2=$
\bea
\nnb&+&{(-2+d)(-4m^2+6dm^2-2d^2m^2-4p^2+5dp^2-d^2p^2+12m^2\xi-10dm^2\xi+2d^2m^2\xi)\over(4(-4+d)(-3+d)m^2p^2}{(\TAIcarre)^2}\\
\nnb&+&{(-1+d)^2\over2(-3+d)}{\TAI}{\TBIm}\\
\nnb&+&{(-16m^2+6dm^2+dp^2)(1-d-3\xi+d\xi)\over2(-4+d)p^2}{\TJIm}\\
\nnb&+&{2m^2(-6m^2+2dm^2-2p^2+dp^2)(1-d-3\xi+d\xi)\over(-4+d)(-3+d)p^2}{\TJIdm}\\
\eea

$F4_2=$
\bea
\nnb&+&{(-2+d)(240m^2p^2-240dm^2p^2-20160m^4\xi)\over1920(-5+d)(-3+d)m^4p^2\xi}{(\TAIcarre)^2}\\
\nnb&+&{(-2+d)(+52832dm^4\xi-37440d^2m^4\xi+9696d^3m^4\xi)\over1920(-5+d)(-3+d)m^4p^2\xi}{(\TAIcarre)^2}\\
\nnb&+&{(-2+d)(-832d^4m^4\xi+1864m^2p^2\xi-4008dm^2p^2\xi)\over1920(-5+d)(-3+d)m^4p^2\xi}{(\TAIcarre)^2}\\
\nnb&+&{(-2+d)(+3352d^2m^2p^2\xi-1208d^3m^2p^2\xi+128d^4m^2p^2\xi)\over1920(-5+d)(-3+d)m^4p^2\xi}{(\TAIcarre)^2}\\
\nnb&+&{(-2+d)(-420p^4\xi+509dp^4\xi-225d^2p^4\xi+43d^3p^4\xi)\over1920(-5+d)(-3+d)m^4p^2\xi}{(\TAIcarre)^2}\\
\nnb&+&{(-2+d)(-3d^4p^4\xi+96320m^4\xi^2-115104dm^4\xi^2)\over1920(-5+d)(-3+d)m^4p^2\xi}{(\TAIcarre)^2}\\
\nnb&+&{(-2+d)(+48768d^2m^4\xi^2-8800d^3m^4\xi^2+576d^4m^4\xi^2)\over1920(-5+d)(-3+d)m^4p^2\xi}{(\TAIcarre)^2}\\
\nnb&+&{(-2+d)(-3128m^2p^2\xi^2+5832dm^2p^2\xi^2-3432d^2m^2p^2\xi^2)\over1920(-5+d)(-3+d)m^4p^2\xi}{(\TAIcarre)^2}\\
\nnb&+&{(-2+d)(+792d^3m^2p^2\xi^2-64d^4m^2p^2\xi^2-60p^4\xi^2)\over1920(-5+d)(-3+d)m^4p^2\xi}{(\TAIcarre)^2}\\
\nnb&+&{(-2+d)(+107dp^4\xi^2-59d^2p^4\xi^2+13d^3p^4\xi^2-d^4p^4\xi^2)\over1920(-5+d)(-3+d)m^4p^2\xi}{(\TAIcarre)^2}\\
\nnb&+&{1680m^4-6840dm^4+6480d^2m^4-1320d^3m^4\over1920m^4}{\TAI}{\TBIo}\\
\nnb&+&{-240m^2p^2+640dm^2p^2-540d^2m^2p^2+140d^3m^2p^2\over1920m^4}{\TAI}{\TBIo}\\
\nnb&+&{56p^4-66dp^4+25d^2p^4-3d^3p^4-16080m^4\xi+23640dm^4\xi\over1920m^4}{\TAI}{\TBIo}\\
\nnb&+&{-8400d^2m^4\xi+840d^3m^4\xi+400m^2p^2\xi-720dm^2p^2\xi\over1920m^4}{\TAI}{\TBIo}\\
\nnb&+&{380d^2m^2p^2\xi-60d^3m^2p^2\xi+8p^4\xi-14dp^4\xi\over1920m^4}{\TAI}{\TBIo}\\
\nnb&+&{7d^2p^4\xi-d^3p^4\xi-6480m^4\xi^2+9720dm^4\xi^2\over1920m^4}{\TAI}{\TBIo}\\
\nnb&+&{-3600d^2m^4\xi^2+360d^3m^4\xi^2-2160m^4\xi^3\over1920m^4}{\TAI}{\TBIo}\\
\nnb&+&{3240dm^4\xi^3-1200d^2m^4\xi^3+120d^3m^4\xi^3\over1920m^4}{\TAI}{\TBIo}\\
\nnb&+&{16128m^4-37024dm^4+18272d^2m^4-2496d^3m^4\over960m^2p^2}{\TJIm}\\
\nnb&+&{-56m^2p^2+784dm^2p^2-1336d^2m^2p^2+384d^3m^2p^2\over960m^2p^2}{\TJIm}\\
\nnb&+&{168p^4-191dp^4+72d^2p^4-9d^3p^4-77056m^4\xi\over960m^2p^2}{\TJIm}\\
\nnb&+&{67040dm^4\xi-18912d^2m^4\xi+1728d^3m^4\xi\over960m^2p^2}{\TJIm}\\
\nnb&+&{-968m^2p^2\xi-1312dm^2p^2\xi+1112d^2m^2p^2\xi\over960m^2p^2}{\TJIm}\\
\nnb&+&{-192d^3m^2p^2\xi+24p^4\xi-41dp^4\xi+20d^2p^4\xi-3d^3p^4\xi\over960m^2p^2}{\TJIm}\\
\nnb&+&{(4m^2+p^2)(-2016m^4+3872dm^4-832d^2m^4)\over960m^2p^2}{\TJIdm}\\
\nnb&+&{(4m^2+p^2)(64m^2p^2-208dm^2p^2+128d^2m^2p^2)\over960m^2p^2}{\TJIdm}\\
\nnb&+&{(4m^2+p^2)(-28p^4+19dp^4-3d^2p^4+9632m^4\xi)\over960m^2p^2}{\TJIdm}\\
\nnb&+&{(4m^2+p^2)(-4768dm^4\xi+576d^2m^4\xi-208m^2p^2\xi)\over960m^2p^2}{\TJIdm}\\
\nnb&+&{(4m^2+p^2)(272dm^2p^2\xi-64d^2m^2p^2\xi)\over960m^2p^2}{\TJIdm}\\
\nnb&+&{(4m^2+p^2)(-4p^4\xi+5dp^4\xi-d^2p^4\xi)\over960m^2p^2}{\TJIdm}\\
\eea

$F5_2=$
\bea
\nnb&-&{416m^2-576dm^2+264d^2m^2-40d^3m^2\over48(-3+d)m^2p^2}{(\TAIcarre)^2}\\
\nnb&-&{+48p^2-53dp^2+18d^2p^2-d^3p^2-608m^2\xi\over48(-3+d)m^2p^2}{(\TAIcarre)^2}\\
\nnb&-&{768dm^2\xi-312d^2m^2\xi+40d^3m^2\xi-6p^2\xi\over48(-3+d)m^2p^2}{(\TAIcarre)^2}\\
\nnb&-&{11dp^2\xi-6d^2p^2\xi+d^3p^2\xi\over48(-3+d)m^2p^2}{(\TAIcarre)^2}\\
\nnb&-&{(-1+d)(4m^2+p^2)(-1+\xi)(-d-4\xi+d\xi)\Gamma[(-1+d)/2]\over32\Gamma(1/2+d/2)}{\TBIo}{\TBIm}\\
\nnb&-&{8(16m^2-8dm^2+3p^2-2dp^2-24m^2\xi)\over32}{\TBIo}{\TBIm}\\
\nnb&-&{8(8dm^2\xi-6p^2\xi+2dp^2\xi)\over32}{\TBIo}{\TBIm}\\
\nnb&+&{3(-1+d)m^2(-1+\xi)(-d-4\xi+d\xi)\Gamma[(-1+d)/2]}{\TAI}\\
\nnb&+&{192m^2-186dm^2+42d^2m^2+26p^2-15dp^2\over48m^2}{\TBIo}\\
\nnb&+&{d^2p^2-264m^2\xi+204dm^2\xi-36d^2m^2\xi\over48m^2}{\TBIo}\\
\nnb&+&{-2p^2\xi+3dp^2\xi-d^2p^2\xi-24m^2\xi^2+30dm^2\xi^2\over48m^2}{\TBIo}\\
\nnb&+&{-6d^2m^2\xi^2\over48m^2}{\TBIo}\\
\nnb&+&{-4m^2+4dm^2-3p^2+2dp^2\over8(-3+d)m^2}{\TBIm}\\
\nnb&-&{p^2(8m^2+p^2)\over8}{\TFI}\\
\nnb&-&{1664m^2-2096dm^2+872d^2m^2-120d^3m^2\over24(-2+d)p^2}{\TJIm}\\
\nnb&-&{288p^2-283dp^2+74d^2p^2-3d^3p^2-2432m^2\xi\over24(-2+d)p^2}{\TJIm}\\
\nnb&-&{2768dm^2\xi-1016d^2m^2\xi+120d^3m^2\xi-114p^2\xi\over24(-2+d)p^2}{\TJIm}\\
\nnb&-&{121dp^2\xi-38d^2p^2\xi+3d^3p^2\xi\over24(-2+d)p^2}{\TJIm}\\
\nnb&-&{2496m^4-3040dm^4+1216d^2m^4\over24(-3+d)(-2+d)p^2}{\TJIdm}\\
\nnb&-&{-160d^3m^4+960m^2p^2-1068dm^2p^2+376d^2m^2p^2\over24(-3+d)(-2+d)p^2}{\TJIdm}\\
\nnb&-&{-44d^3m^2p^2+96p^4-83dp^4+18d^2p^4-d^3p^4\over24(-3+d)(-2+d)p^2}{\TJIdm}\\
\nnb&-&{-3648m^4\xi+4000dm^4\xi-1408d^2m^4\xi+160d^3m^4\xi\over24(-3+d)(-2+d)p^2}{\TJIdm}\\
\nnb&-&{-936m^2p^2\xi+1044dm^2p^2\xi-376d^2m^2p^2\xi\over24(-3+d)(-2+d)p^2}{\TJIdm}\\
\nnb&-&{44d^3m^2p^2\xi-6p^4\xi+11dp^4\xi-6d^2p^4\xi+d^3p^4\xi\over24(-3+d)(-2+d)p^2}{\TJIdm}\\
\eea

$F6_2=$
\bea
\nnb&{}&{(-2+d)(-1+d)d\over4(-3+d)m^2}{(\TAIcarre)^2}\\
\eea

$F7_2=$
\bea
\nnb&{}&{3(-1+d+\xi)\over4}{\TJIm}\\
\eea

$F9_2=$
\bea
\nnb&{}& {(-2+d)(-1+d)(1-\xi+d\xi)\over8(-5+d)(-3+d)m^2\xi}{(\TAIcarre)^2}\\
\eea

$F10_2=$
\bea
\nnb&-&{(-1+d)(-14+43d-11d^2+134\xi-63d\xi)\over16}{\TAI}{\TBIo}\\
\nnb&-&{(-1+d)(7d^2\xi+54\xi^2-27d\xi^2)\over16}{\TAI}{\TBIo}\\
\nnb&-&{(-1+d)(3d^2\xi^2+18\xi^3-9d\xi^3+d^2\xi^3)\over16}{\TAI}{\TBIo}\\
\eea

\end{document}